\documentclass[showpacs,prb,aps]{revtex4}
\begin{document} 
\title{Recombination in Semiconductors: Appearance of Non-equilibrium Carriers
due to Injection or Redistribution in the Sample}
\author{I.~N.~Volovichev}
\altaffiliation{Permanent address: Institute for
Radiophysics and Electronics, National Academy of Sciences of
Ukraine, Kharkov 310085, Kharkov 310085, Ukraine.}
\author{G.~Espejo}
\email{gespejo@fis.cinvestav.mx}
\author{Yu.~G.~Gurevich}
\email{gurevich@fis.cinvestav.mx}
\affiliation{Departamento de F\'{\i}sica, CINVESTAV---I.P.N., \\
Apdo.~Postal 14--740, M\'{e}xico, D.F.~07000, M\'{e}xico}
\author{O.~Yu.~Titov}
\email{oleg.titov@aleph-tec.com}
\affiliation{CICATA---I.P.N., Jos\'{e} Siurob No.~10, \\
Col.~Alameda, C.P.~76040, Santiago de Quer\'{e}taro, Qro., M\'{e}xico}
\author{A.\ Meriuts}
\affiliation{Kharkov Polytechnic University, 21 Frunze Str., \\
Kharkov 310002, Ukraine}
\date{\today} 
\draft
\begin{abstract} 
It is shown that the traditional approach for
consideration of recombination under condition of steady-state current in
the absence of external carrier generation is internally contradictory.
Sometimes the approach leads to obviously incorrect results. Such
situations are demonstrated and a new method for consideration of
recombination is proposed. 
\end{abstract} 
\pacs{72.20.Jv}
\maketitle

In the present paper we would like to consider some internal
inconsistencies in the conventional description when we have 
carrier recombination under condition of steady-state current.
Practically all text-books\cite{Newm,Shokley,Bonch,KS,Zi} 
present technique to solve
the given problem based on the solution of a set of continuity
equations:
\[
{\rm div}{\bf j}_{n} =
e R_{n},\qquad 
{\rm div}{\bf j}_{p} = -e R_{p}
\]
where $n$ and $p$ are the electron and hole concentration, ${\bf
j}_{n,p}$ are the current of electrons and holes, $R_{n,p}$ are the
electron and hole recombination rate respectively. Here we
consider the absencens of external generation of carriers (by
light, etc.). Thus the nonequilibrium carriers are a result
of injection or accumulation of carriers near potential barriers
at interfaces. For small concentration of non-equilibrium carriers
$\delta n \equiv n - n_{0} \ll n_{0}$, 
$\delta p \equiv p - p_{0} \ll p_{0}$
(where $n_{0}$, $p_{0}$ are concentration of electrons and holes
without the current) the recombination rates are widely assumed to
be of the following form:
\begin{equation}
\label{main}
R_{n} = \frac{\delta n}{\tau_{n}},
\qquad 
R_{p} = \frac{\delta p}{\tau_{p}},
\end{equation}
where $\tau_{n}$, $\tau_{p}$ are life times of electrons and holes
respectively.

Obviously, by virtue of charge preservation law an extra
condition
\begin{equation}
\label{equality}
R_{n} = R_{p}
\end{equation}
should hold. Although it is not mentioned in the literature that
Eq.~(\ref{equality}) makes the system overdetermined. Some 
authors\cite{Bonch} use the latter expression as
an equation to find the carrier concentration or to reduce by one
the number of unknowns in the problem. However, such approach
seems to be incorrect because the Eq.~(\ref{equality}) is not a
new condition for the concentration of non-equilibrium carriers,
rather it is the criterion for correctness of the recombination
description, and should fulfill identically at any concentration
of non-equilibrium carriers. Probably due to this reason other
approach frequently is used,\cite{Newm,Shokley,Bonch,KS,Peka,Zi,Lamp} 
assuming
\begin{equation}
\label{trad}
R_{n} \equiv R_{p} = \frac{\delta p}{\tau_{p}},
\end{equation}
where $\delta p$ means the non-equilibrium 
concentration of minority carriers.
Just this approach is classical and is used widely both in
text-books, and in papers devoted to kinetic phenomena in
semiconductors.

This description
possesses a serious inconsistency. It becomes especially obvious
if we consider injection of majority carriers. From physical reasons it
is evident: injected non-equilibrium majority carriers should recombine.
While from a formal 
point of view, as far as non-equilibrium minority carriers do not
occur ($\delta p=0$) the recombination rates are also equal to
zero $R_n=R_p=0$. Below we consider more cases, when the
conventional description of a recombination is unacceptable.

Elimination of the given inconsistency is not
complicated and can be based on the well known formulas
presented in the above mentioned text-books. 
It is easy to get the following expressions for
recombination processes\cite{Bonch}
by a sequential
statistical consideration of transitions between valence and
conduction bands and between the bands and impurity levels (if
they are present):
\begin{equation}
\label{interband}
R_{n} = R_{p} = \alpha(np - n^{2}_{i})
\end{equation}
for interband recombination and
\begin{equation}
\label{Read} 
R_{n} = R_{p} = 
\frac{np - n^{2}_{i}}{\tau_{n}^{0}(p + p_{1}) + \tau_{p}^{0}(n + n_{1})}
\end{equation}
for recombination through an impurity level according to the
Shockley-Read model.\cite{Srm}

Here $\alpha$ is the recombination factor, $n_{i}$ is the carrier
concentration of the intrinsic semiconductor, $\tau_{n,p}^{0}$, $n_{1}$,
and $p_{1}$ are characteristics of the impurity level.

We should emphasize once again that the equality $R_{n} = R_{p}$ is
inherent in correct model and holds identically, not causing
overdetermination of the system.

Linearizing these expressions for the case of weak deviation from
the condition of thermodynamic equilibrium (the changes of carrier
concentrations due to current are small on comparison with their
equilibrium values), we have for both cases:
\begin{equation}
\label{rates}
R_{n} = R_{p} = 
\frac{\delta n}{\tau_{n}} + \frac{\delta p}{\tau_{p}},\quad
\text{with } \frac{\tau_{n}}{\tau_{p}}=\frac{n_{0}}{p_{0}}.
\end{equation}

Just these expressions, being used in Eqs.~(\ref{main}), eliminate
all mentioned above inconsistencies.

Also let us pay attention to another aspects of the problem.
As it is easy to notice from Eqs.~(\ref{rates}), in bipolar
semiconductors generally it is impossible to introduce correctly a
life time of carriers concept. An exception is the case of
quasineutrality\cite{QN} when $\delta
n=\delta p$ and the introduction of unified life time of carriers
$\tau^{-1} \equiv \tau_{n}^{-1} + \tau_{p}^{-1}$ becomes possible. In the
absence of quasineutrality it is plausible to speak about life
time of carriers and to use the conventional approach
Eq.~(\ref{trad}) only if $(\delta n / \delta p) \ll
(\tau_n / \tau_p) = (n_0/p_0)$ (here, as it was mentioned above, $n$ means
concentration of majority carriers). Naturally, the fulfilment of
this condition is not {\it a priori} obvious.

Let us notice that another baseless idea is widespread, namely
presence of only interband recombination is sufficient condition
for the equality $\delta n=\delta p$ to be
fulfilled.\cite{Newm,Bonch,KS} But there is no proof for this
conclusion, and moreover, a case of injection obviously
contradicts it.

Thus, in general case for static current either it is impossible
to introduce correctly the concept of life time or life times are
the same for both electrons and holes.

For the sake of justice we should notice, that as a rule in the
literature either injection phenomena\cite{Lamp}
(when linearized equations are not valid due to high concentration of non-equilibrium carriers) or just the quasineutrality
approach are considered. From our point of view, it is just the
only explanation why so internally contradictory method to
describe kinetic phenomena with the presence of
recombination is widely used in many text-books and monographies.

Similar problem arises for description of a surface
recombination. Correct expression for it should be obtained from
consideration of transitions between the bands and impurity levels
both inside the semiconductor, and between bands of different
semiconductors (or between bands of the semiconductor and metal in
case of a metal-semiconductor contact). 
The correct expression for the surface recombination
rate should ensure identical equality of electron and hole
recombination rates at any concentration of non-equilibrium
carriers at the contact
(as in case of volume recombination).
It is obvious that in general case the
expression of the following form meets the requirement:
\begin{equation}
S =s_{n} \delta n + s_{p} \delta p,
\end{equation}
where coefficients $s_{n,p}$ should be obtained from microscopic
consideration of corresponding transitions.

Some authors use that formula at initial stage of calculations,
then giving it up and proceeding to the traditional (but
incorrect!) expression $S_{n} = s_{n}\delta n$, 
$S_{p} = s_{p}\delta p$. 
Rather characteristic consideration is presented in
Ref.~\onlinecite{Peka}, where quasilevels of electrons and holes are
assumed to coincide. That, in its turn, implicitly presumes
infinite surface (or volume) recombination rate.

The situation becomes more complicated, if there is an
inhomogeneous temperature distribution (temperature field) in the
semiconductor.
It is problematic to choose a concentration which
can be considered as a level of reference of non-equilibrium
concentration (for example, which one should be taken as the
reference: $n_{0}(T_0)$, $n_{0}[T(x)]$, $n_{0}[\overline{T}(x)]$, etc).
A way out from this position can be found by successive
consideration of process of establishment of ``equilibrium'' (or in
constancy of electrochemical potential in the sample) in the
circuit.\cite{Gur} Evidently, this method is
acceptable only if the temperature field is given. Other,
universal and more simple, from our point of view, way to solve
this problem is to return to statistical consideration of
transitions of carriers between bands and impurity levels in the
presence of temperature field. In this case the value of
concentration in thermodynamic equilibrium (in the absence of the
temperature gradient) or the values corresponding to the mean
temperature of the sample can be accepted as the reference level
of concentration (i.e. $n_{0}$ and $p_{0}$). This method works as well
in the case, when the temperature field should be determined
self-consistently.

Thus for interband recombination we come back to expressions
(\ref{interband}), where the values $\alpha$ and $n_{i}$ are
functions of temperature ($\alpha=\alpha[T(x)]$, $n_{i} = n_{i}[T(x)]$).
For small deviation from the equilibrium state (after linearization)
we receive the following expressions for volume recombination rate
in the temperature field:
\begin{equation}
\label{gamma}
R_{n} = R_{p} =
\frac{\delta n}{\tau_{n}} + \frac{\delta p}{\tau_{p}} + \gamma\delta T,
\end{equation}
where $\gamma\equiv ({1}/{2\tau})({\partial n_{i}}/{\partial
T})$; $\delta T \equiv T - T_{0}$; $T_{0}$ is the equilibrium temperature
(above-mentioned reference level of temperature). Let us note that
in the approximation of small nonequilibrium carrier
concentrations ($\delta n \ll n_{0}$, $\delta p \ll p_{0}$) thermal
dependence of the recombination factor $\alpha$ does not manifest
itself.

Thus, the presence of a temperature gradient results in appearance
of an additional term in expressions for recombination rates. This
term takes into account the change of the rate of thermal
generation (which, as it is well known, is proportional to squared
concentration of intrinsic semiconductor at given temperature). As
far as we know, anywhere but Ref.~\onlinecite{Gur}, this contribution has
not been taken into account in consideration of thermoelectric
phenomena.

Finally, if there is no unified
temperature of carriers and phonons (for example,  hot
electrons situation\cite{Bonch}), it is easy to prove by a
similar way that one more term appears in the expressions for
recombination rates.It proportional to the heating of 
carriers:\cite{PRB99}
\begin{equation}
\label{general}
R_{n} = R_{p} =
\frac{\delta n}{\tau_{n}} + 
\frac{\delta p}{\tau_{p}} + 
\gamma(T(x) - T_{0}) + \beta(T_{e} - T_{0}),
\end{equation}
where $T_{e}$ is the temperature of hot electrons, $T(x)$ is assumed
to be the lattice temperature. The coefficient $\beta$ depends on
concentration of carriers, the lattice temperature and capture
factor of carriers by the impurity level (or corresponding value
for interband recombination).

Let us note that a coordinate dependence of electron temperature
$T_{e} = T_{e}(x)$, that is a general case, does not affect the 
term $\gamma(T(x) - T_{0})$
from Eq.~(\ref{general}), which is the
same as in Eq.~(\ref{gamma}). This is due to the fact that this term
describes thermal generation of carriers, which can not be
influenced by population of the conduction band (i.e. by electron
temperature) just because of huge concentration of free states in
the conduction band.

Thus, we would like to pay attention of researchers to the fact,
that holding a firm place approach to the description of a
recombination in steady-state mode is internally contradictory and
in many cases, especially common to modern problems,
its hasty application can lead to incorrect results.
We would like to mention, as an example, thin-film
devices, where it is easy to disturb the quasineutrality and to
generate essential gradients of temperature and energy
nonequilibrium of carriers due to carriers heating by an applied field.

This work has been partially supported by CONACyT---M\'{e}xico.


\begin{references}
\bibitem{Newm} 
D.~A.~Neamen, 
{\it Semiconductor Physics and Devices: Basic Principles} 
(Richard D. Irwin, Boston, 1992).

\bibitem{Shokley} 
W.~Shockley, 
{\it Electrons and holes in Semiconductor with Applications to Transistor
Electronics}
(Toronto-New York-London, 1950).

\bibitem{Bonch} 
V.~L.~Bonch-Bruevich and S.~G.~Kalashnikov,
{\it Physics of the Semiconductors} 
(VEB Deutscher Verlag der Wissenschaften, Berlin, 1982). 
[In German]

\bibitem{KS} 
K.~Seeger, 
{\it Semiconductor Physics: An Introduction
(6th Ed.)} (Springer Verlag, New York, 1997).

\bibitem{Zi} 
S.~M.~Sze, 
{\it Physics of Semiconductor Devices}
(John Wiley \& Sons, New York,  1981).

\bibitem{Peka} 
G.~P.~Peka, 
{\it Physics Phenomena at Semiconductor Surfaces} 
(Vyscha Shkola, Kiev, 1984). 
[In Russian]

\bibitem{Lamp} 
M.~Lamport and P.~Mark, 
{\it Currents Injection in Solids}
(Academic Press, New York, 1970).

\bibitem{Srm} 
W.~Shockley and W.~T.~Read, 
Phys. Rev. {\bf 87} 835 (1952).

\bibitem{QN} 
V.~P.~Silin and A.~A.~Rukhadze, 
{\it Electromagnetic Properties of the Plasma and Plasma Related Media}
(Atomizdat, Moskow, 1961).
[In Russian]

\bibitem{Gur} 
Yu.~G.~Gurevich, G.~N.~Logvinov, O.~I.~Lubimov and
O.~Yu.~Titov, 
Phys.~Rev.~B {\bf 51} 6999 (1995).

\bibitem{PRB99} 
Yu.~G.~Gurevich, I.~N.~Volovichev, 
Phys.~Rev.~B {\bf 60} 7715 (1999).

\end{references}
\end{document}